


\font\titlefont = cmr10 scaled\magstep 4
\font\sectionfont = cmr10
\font\littlefont = cmr5
\font\eightrm = cmr8

\def\ss{\scriptstyle}
\def\sss{\scriptscriptstyle}

\newcount\tcflag
\tcflag = 0  

\ifnum\tcflag = 0 \magnification = 1200 \fi  

\global\baselineskip = 1.2\baselineskip
\global\parskip = 4pt plus 0.3pt
\global\abovedisplayskip = 18pt plus3pt minus9pt
\global\belowdisplayskip = 18pt plus3pt minus9pt
\global\abovedisplayshortskip = 6pt plus3pt
\global\belowdisplayshortskip = 6pt plus3pt

\def\barsoff{\overfullrule=0pt}


\def\endignore{}
\def\ignore #1\endignore{}

\newcount\dflag
\dflag = 0


\def\monthname{\ifcase\month
\or January \or February \or March \or April \or May \or June%
\or July \or August \or September \or October \or November %
\or December
\fi}

\newcount\dummy
\newcount\minute  
\newcount\hour
\newcount\localtime
\newcount\localday
\localtime = \time
\localday = \day

\def\advanceclock#1#2{ 
\dummy = #1
\multiply\dummy by 60
\advance\dummy by #2
\advance\localtime by \dummy
\ifnum\localtime > 1440 
\advance\localtime by -1440
\advance\localday by 1
\fi}

\def\settime{{\dummy = \localtime%
\divide\dummy by 60%
\hour = \dummy
\minute = \localtime%
\multiply\dummy by 60%
\advance\minute by -\dummy
\ifnum\minute < 10
\xdef\spacer{0} 
\else \xdef\spacer{}
\fi %
\ifnum\hour < 12
\xdef\ampm{a.m.} 
\else
\xdef\ampm{p.m.} 
\advance\hour by -12 %
\fi %
\ifnum\hour = 0 \hour = 12 \fi
\xdef\timestring{\number\hour : \spacer \number\minute%
\thinspace \ampm}}}



\def\endtitle{}
\def\title#1\endtitle{\vskip.5in\titlefont
\global\baselineskip = 2\baselineskip
#1\vskip.4in
\baselineskip = 0.5\baselineskip\rm}

\def\endauthors{}
\def\authors#1\endauthors{#1}

\def\endabstract{}
\def\abstract#1\endabstract{\vskip .3in%
\centerline{\sectionfont\bf Abstract}%
\vskip .1in
\noindent#1}

\def\nopageonenumber{\footline={\ifnum\pageno<2\hfil\else
\hss\tenrm\folio\hss\fi}}  

\newcount\nsection
\newcount\nsubsection

\def\section#1{\global\advance\nsection by 1
\nsubsection=0
\bigskip\noindent\centerline{\sectionfont \bf \number\nsection.\ #1}
\bigskip\rm\nobreak}

\def\subsection#1{\global\advance\nsubsection by 1
\bigskip\noindent\sectionfont \sl \number\nsection.\number\nsubsection)\
#1\bigskip\rm\nobreak}

\def\topic#1{{\medskip\noindent $\bullet$ \it #1:}}
\def\endtopic{\medskip}

\def\appendix#1#2{\bigskip\noindent%
\centerline{\sectionfont \bf Appendix #1.\ #2}
\bigskip\rm\nobreak}


\newcount\nref
\global\nref = 1

\def\therefs{}


\def\ref#1#2{\xdef #1{[\number\nref]}
\ifnum\nref = 1\global\xdef\therefs{\item{[\number\nref]} #2\ }
\else
\global\xdef\oldrefs{\therefs}
\global\xdef\therefs{\oldrefs\vskip.1in\item{[\number\nref]} #2\ }%
\fi%
\global\advance\nref by 1
}

\def\listrefs{\vfill\eject\section{References}\therefs}


\newcount\nfoot
\global\nfoot = 1

\def\foot#1#2{\xdef #1{(\number\nfoot)}
\footnote{${}^{\number\nfoot}$}{\eightrm #2}
\global\advance\nfoot by 1
}


\newcount\nfig
\global\nfig = 1
\def\thefigs{} 

\def\figure#1#2{\xdef #1{(\number\nfig)}
\ifnum\nfig = 1\global\xdef\thefigs{\item{(\number\nfig)} #2\ }
\else
\global\xdef\oldfigs{\thefigs}
\global\xdef\thefigs{\oldfigs\vskip.1in\item{(\number\nfig)} #2\ }%
\fi%
\global\advance\nfig by 1 } 

\def\figurecaptions{\vfill\eject\section{Figure Captions}\thefigs}

\def\fig#1{\xdef #1{(\number\nfig)}
\global\advance\nfig by 1 } 


\newcount\cflag
\newcount\nequation
\global\nequation = 1
\def\eqlabel{(1)}

\def\nexteqno{\ifnum\cflag = 0
\global\advance\nequation by 1
\fi
\global\cflag = 0
\xdef\eqlabel{(\number\nequation)}}

\def\lasteqno{\global\advance\nequation by -1
\xdef\eqlabel{(\number\nequation)}}

\def\label#1{\xdef #1{(\number\nequation)}
\ifnum\dflag = 1
{\escapechar = -1
\xdef\draftname{\littlefont\string#1}}
\fi}

\def\clabel#1#2{\xdef\eqlabel{(\number\nequation #2)}
\global\cflag = 1
\xdef #1{\eqlabel}
\ifnum\dflag = 1
{\escapechar = -1
\xdef\draftname{\string#1}}
\fi}

\def\cclabel#1#2{\xdef\eqlabel{#2)}
\global\cflag = 1
\xdef #1{\eqlabel}
\ifnum\dflag = 1
{\escapechar = -1
\xdef\draftname{\string#1}}
\fi}


\def\eeq{}

\def\eqnn #1\eeq{$$ #1 $$}

\def\eq #1\eeq{
\ifnum\dflag = 0
{\xdef\draftname{\ }}
\fi 
$$ #1
\eqno{\eqlabel \rlap{\ \draftname}} $$
\nexteqno}



\def\eol{& \eqlabel \rlap{\ \draftname} \crcr
\nexteqno
\xdef\draftname{\ }}

\def\eeol{& \eqlabel \rlap{\ \draftname}
\nexteqno
\xdef\draftname{\ }}



\def\eqa #1\eeq{
\ifnum\dflag = 0
{\xdef\draftname{\ }}
\fi 
$$ \eqalignno{ #1 } $$
\global\cflag = 0}


\def\ie{{\it i.e.\/}}

\def\etc{{\it etc.\/}}
\def\etal{{\it et.al.\/}}


\def\npb#1#2#3{{\it Nucl.\ Phys.} {\bf B#1} (19#2) #3}
\def\plb#1#2#3{{\it Phys.\ Lett.} {\bf #1B} (19#2) #3}

\def\prd#1#2#3{{\it Phys.\ Rev.} {\bf D#1} (19#2) #3}

\def\prl#1#2#3{{\it Phys.\ Rev.\ Lett.} {\bf #1} (19#2) #3}


\global\nulldelimiterspace = 0pt



\def\frac#1#2{{{#1} \over {#2}}\,}  
\def\hf{{1\over 2}}
\def\nth#1{{1\over #1}}


\def\Dsl{\hbox{/\kern-.6700em\it D}} 
\def\dsl{\hbox{/\kern-.5300em$\partial$}}
\def\pxpsl{\hbox{/\kern-.5600em$p$}}
\def\ssl{\hbox{/\kern-.5300em$s$}}
\def\epssl{\hbox{/\kern-.5100em$\epsilon$}}
\def\delsl{\hbox{/\kern-.6300em$\nabla$}}
\def\lxpsl{\hbox{/\kern-.4300em$l$}}
\def\elxpsl{\hbox{/\kern-.4500em$\ell$}}
\def\kxpsl{\hbox{/\kern-.5100em$k$}}
\def\qxpsl{\hbox{/\kern-.5000em$q$}}
\def\sla#1{\raise.15ex\hbox{$/$}\kern-.57em #1}



\def\roughly#1{\mathrel{\raise.3ex
\hbox{$#1$\kern-.75em\lower1ex\hbox{$\sim$}}}}






\def\ssl{{\sss L}}

\def\ssy{{\sss Y}}







\def\GeV{{\rm \ GeV}}


\nopageonenumber
\baselineskip = 18pt
\barsoff


\def\ss{\scriptstyle}

\def\sutwo{$SU_\ssl(2)$}
\def\suthree{$SU_c(3)$}

\def\gwk{$SU_\lft(2) \times U_Y(1)$}

\def\mw{M_{\sss W}}
\def\mz{M_{\sss Z}}
\def\mt{m_t}
\def\mh{m_{\sss H}}

\def\rht{{\sss R}}
\def\lft{{\sss L}}
\def\sw{s_w}
\def\cw{c_w}

\def\alr{A_{\lft\rht}}
\def\aeptau{A_e(P_\tau)}
\def\bfone{{\bf 1}}
\def\bftwo{{\bf 2}}
\def\bfthree{{\bf 3}}


\line{hep-ph/9407203 \hfil McGill-94/27, NEIP-94-005}
\rightline{June, 1994.}
\vskip 0.05in

\title
\centerline{Negative S and Light New Physics*}
\endtitle

\footnote{}{${}^*$ {\eightrm Research supported by the Swiss National
Foundation.}}
\vskip 0.15in
\authors
\centerline{P. Bamert${}^a$ and C.P. Burgess${}^{a,b}$}
\vskip .15in
\centerline{\it ${}^a$ Institut de Physique, Universit\'e de
Neuch\^atel, CH-2000 Neuch\^atel, Switzerland.}
\vskip .1in
\centerline{\it ${}^b$ Physics Department, McGill University}
\centerline{\it 3600 University St., Montr\'eal, Qu\'ebec, CANADA, H3A 2T8.}
\endauthors

\vskip .1in

\abstract
The several sigma difference between SLD's recent precise measurement of $\alr$
and the corresponding LEP results tends to reinforce the earlier trends in the
data towards negative values for the Peskin-Takeuchi parameter $S$ and $T$.
Motivated by this not yet statistically significant, but suggestive, trend, we
explore which kinds of new particles can (1) contribute dominantly to new
physics through oblique corrections, (2) produce negative values for $S$ and
$T$, and (3) not be in conflict with any other experiments, on or off the $Z$
resonance. We are typically led to models which involve new particles with
masses that are not much heavier than $\mz/2$, and so which would
also have implications for other experiments in the near future. We show how
the
analysis of such `light-new-physics' models in terms of oblique parameters
requires
the interpretation of the data in terms of modified parameters, $S'$ and $T'$,
whose difference from $S$ and $T$ improves the available parameter space of the
models.
\endabstract


\vfill\eject

\section{Introduction}

\ref\moriond{B. Pietrzyk, in the proceedings of the XXIXth Rencontres de
Moriond, March 1994, preprint LAPP-EXP-94.07 (hep-ex/9406001).}
\ref\sld{The SLD Collaboration, F. Abe \etal, preprint SLAC-PUB-6456,
to appear in {\it Physical Review Letters};
M. Woods, {\it ibid},  preprint SLAC-PUB-6493.}
\ref\comparison{J. Erler, preprint UPR-0619T (hep-ph/9406429).}
\ref\lps{B. Lynn, M. Peskin, R. Stuart, in {\it Physics at LEP} vol. 1, CERN
Report 86-02, pp. 90 -- 152.}
\ref\peskt{M.E. Peskin and T. Takeuchi, \prl{65}{90}{964};
\prd{46}{92}{381}.}
\ref\peskin{M. Peskin, as quoted in Ref. [8] below.}

As the accuracy of the measurements of the properties of the the $Z$-boson
resonance continue to improve, the tests of the standard electroweak theory are
reaching new levels of precision \moriond. Recently the numerous measurements
at
CERN's Large Electron Positron ring have been joined by the measurement of the
left-right asymmetry, $\alr$, at the Stanford Linear Collider \sld. This last
measurement --- touted as the most accurate single electroweak measurement on
the $Z$ resonance --- implies an electron-$Z$ coupling which lies some
2.5-$\sigma$ away from the corresponding combined LEP values for the same
number. The difference between this coupling as measured by $\alr$, and by the
$\tau$-polarization asymmetry, $\aeptau$, at LEP is over 3 $\sigma$ \moriond,
\comparison.
When analyzed in terms of new-physics contributions to the vector boson vacuum
polarizations \lps\ --- \ie\ in terms of the oblique parameters, $S, T$ and $U$
\peskt\ --- the $\alr$ result tends to push $S$ to more negative values. (A
recent fit \peskin, for instance, gives $S = -0.58 \pm 0.30$ and $T = -0.38 \pm
0.34$, for $\mt = 165 \; \GeV$ and $\mh = 300 \; \GeV$. For $m_t = 174$ GeV,
on the other hand, $S$ does not change appreciably but the central value
for $T$ decreases by $\simeq 0.2$.)
Although this does not yet represent a statistically significant deviation from
the Standard Model (SM), it remains tantalizing that the earlier trend toward
more negative central values for $S$ is reinforced by the newer results.

\ref\negatives{H. Georgi, \npb{363}{91}{301};
E. Gates and J. Terning, \prl{67}{91}{1840};
E. Ma and P. Roy, \prl{68}{92}{2879};
M. Luty and R. Sundrum, \prl{70}{93}{529};
L. Lavoura, L.-F. Li, \prd{48}{93}{234}.}
\ref\newgaugebosons{T.G. Rizzo, preprint SLAC-PUB-6455,
to appear in {\it Phys. Rev. D}.}

With such spiffy new experimental numbers a theorist's fancy inevitably turns
to thoughts of interpretation. A negative value for $S$ is particularly
interesting in this regard, since this was found to be reasonably difficult to
obtain within the context of technicolour models
\negatives. (A recent discussion in terms of new gauge bosons,
motivated by the SLD  $\alr$ measurement, can be found in Ref.
\newgaugebosons.)

\ref\alphabet{I. Maksymyk, C.P. Burgess and D. London, \prd{}{94}{} (in
press); C.P Burgess, S. Godfrey, H. K\"onig, D. London and I. Maksymyk,
\plb{326}{94}{276}.}

In this note we construct several types of models which contribute to precision
electroweak measurements dominantly (or, for some models, exclusively) through
oblique corrections. Furthermore, they do so {\it as if} $S$ were negative. The
qualification `as if' is required because the quantities which appear in the
expressions for the observables are, in general, not $S, T$ and $U$ as they are
usually defined. The basic point is that if the oblique new physics
should not be heavy in comparison to the $W$ and $Z$ masses, then its
contribution to all of the low-energy neutral-current data generally requires
two parameters --- $V$ and $X$ of Ref. \alphabet\ ---  in addition to the usual
three. (The third new quantity, $W$, of Ref. \alphabet\ contributes only to the
$W$-boson width and is ignored here.)

\ref\epsilondefs{G. Altarelli and R. Barbieri, \plb{253}{91}{161};
G. Altarelli, R. Barbieri and S. Jadach, \npb{369}{92}{3}, (erratum)
{\it ibid.}{\bf B376} (1992) 444; G. Altarelli, R. Barbieri and
F. Caravaglios, \npb{405}{93}{3}.}

The necessity for additional parameters might come, at first thought, as a
surprise since oblique contributions to $Z$-pole physics and $\mw$ only involve
three independent observables. One might therefore expect to be able to choose
these to be the standard quantities, $S, T$ and $U$, provided one chose to
work exclusively with $\mw$ and measurements on the $Z$ resonance. This turns
out not to be true.\foot\conversations{We thank David London, Ivan
Maksymyk and Probir Roy for useful conversations on this point.} (The same
objection does not apply to the $\epsilon$ formalism of Ref. \epsilondefs,
but {\it only} if $\mw$ and $Z$-pole data are all that are considered.) It {\it
is}, of
course, true that, for $Z$-pole physics and $\mw$ only, all oblique corrections
to observables can be summarized into three independent quantities.
It is convenient to choose these to be:
\label\primedstu
\eqa
S' &= S + 4 \sw^2 \cw^2 \; V + 4 (\cw^2 - \sw^2) \; X ,\eol
T' &= T + V ,\eol
U' &= U - 4 \sw^2 \cw^2 V + 8 \sw^2 X , \eeol
\eeq
where $\sw$ and $\cw$ denote the sin and cosine of the weak mixing angle,
$\theta_w$.

With this choice $S', T'$ and $U'$ have two very convenient properties. First,
they reduce to $S, T$ and $U$ in the limit of heavy new physics, since in this
case $V$ and $X$ both become completely negligible. Second, these definitions
ensure that all $Z$-pole observables (and $\mw$) depend on $S', T'$
(and $U'$) in precisely the same manner as they do on $S, T$ and $U$ in
the usual analyses. As a consequence, the results of any fits to the present
LEP and SLD data, together with the $W$ mass, apply
verbatim to $S', T'$ and $U'$. The `trend' of the data can therefore be more
properly phrased as a trend towards negative values for $S'$ (and $T'$), with
the fit of Ref. \peskin\ quantitatively implying $S' = -0.58 \pm 0.30$ and $T'
=
-0.38 \pm 0.34$ for the given values of $\mt$ and $\mh$. $S$ and $T$ are
themselves only constrained independently of this by the additional data at
$q^2 \simeq 0$.

This difference between the definitions of the primed and unprimed parameters
is not merely an academic point. We find in our search for models that it
frequently happens that $S'$ and $T'$ take their most negative values when the
new physics is light enough to permit positive, or slightly negative, values
for
$S$ and $T$ to be compensated in $S'$ and $T'$ by negative contributions to $V$
and $X$.  Since the models to which we are led therefore typically involve
comparatively light particles, they can be expected to have more direct
experimental implications in experiments in the comparatively near future.

\section{Models}

We now turn to the construction of models. Our purpose is to survey the
parameter space of simple models to find those for which $S'$ and $T'$ are both
negative,  and are both roughly the same size. We take a conservative approach
and  simply supplement the SM by a few  additional spin-zero or spin-half
particle  types, and explore the one-loop oblique parameters they generate as a
function  of the assumed quantum numbers and masses of the new particles. Of
particular interest among our results are some particular cases of new
particles, since these arise quite  naturally among the low-energy spectrum of
more complicated, but theoretically  better motivated, models (such as the
supersymmetric standard model \etc).

\midinsert

$$\vbox{\tabskip=0pt \offinterlineskip
\halign to \hsize{\strut#& #\tabskip 1em plus 2em minus .5em
&\hfil#\hfil &#&\hfil#\hfil &#&\hfil#\hfil &#&\hfil#\hfil
&#&\hfil#\hfil &#&\hfil#\hfil &#&\hfil#\hfil &#& \hfil#\hfil
&# \tabskip=0pt\cr
\noalign{\hrule}\noalign{\smallskip}\noalign{\hrule}\noalign{\medskip}
&& Multiplet && Optimal Masses && $S'$ && $T'$ && $U'$ && $S$
&& $T$ && $U$ &\cr
&& $(3 \times 2 \times 1)$ && (GeV) &&  &&  &&  &&  &&  && &\cr
\noalign{\medskip}\noalign{\hrule}\noalign{\medskip}
&& (\bfone,\bfone,$Y=1$) && $m = 50$ && -0.01 && -0.006 && 0.002 &&
-0.003 && 0 && 0.003 &\cr
&&  &&  &&  && && &&  &&  && &\cr
&& (\bfone,\bftwo,$Y=\hf$) && $\pmatrix{m_1 \cr m_0 \cr} = \pmatrix{62 \cr
50 \cr}$ && -0.04 && -0.03  && 0.03 && -0.03 && 0.003 && 0.003 &\cr
\noalign{\smallskip}
&& (\bfone,\bftwo,$Y=0$) && $\pmatrix{m_{+\hf} \cr  m_{-\hf} \cr} =
\pmatrix{50 \cr 72.5 \cr}$  && -0.02 && -0.01 && 0.02 && -0.01 &&
0.009 && 0.003 &\cr
\noalign{\smallskip}
&& (\bfone,\bftwo,$Y={3\over 2}$) && $\pmatrix{m_2 \cr m_1 \cr} =
\pmatrix{51 \cr 50 \cr}$ && -0.09 && -0.06 && 0.04 && -0.03 &&
0.00002 && 0.01 &\cr
&&  &&  &&  && && &&  &&  && &\cr
&& (\bfone,\bfthree,$Y=0$)${}^\dagger$ && $\pmatrix{m_1 \cr m_0 \cr m_{-1}
\cr} = \pmatrix{50 \cr 78 \cr 50 \cr}$ && -0.06 && -0.04 && 0.1 && -0.03
&& 0.03 && 0.07 &\cr
\noalign{\smallskip}
&& (\bfone,\bfthree,$Y=1$) && $\pmatrix{m_2 \cr m_1 \cr m_0 \cr} =
\pmatrix{63 \cr 57 \cr 50\cr}$ && -0.2 && -0.1 && 0.1 && -0.1 && 0.003 &&
0.01 &\cr
\noalign{\medskip}\noalign{\hrule}\noalign{\smallskip}\noalign{\hrule} }}$$
${}^\dagger$ {\eightrm Self-conjugate multiplet.}
\smallskip

\centerline{\bf Table I: Exotic Scalars}
\medskip
\noindent {\eightrm One-loop oblique electroweak parameters due to exotic
scalar multiplets. This table displays the masses which `optimize' the
oblique electroweak parameters in the sense described in the text, together
with the resulting optimal values. $\ss ({\bf r_1}, {\bf r_2}, Y=y)$
denotes  the representation of $\ss SU_c(3) \times SU_\ssl (2) \times U_\ssy
(1)$
in which the scalars transform, and $m_q$ represents the mass of a state
having electric charge $q$. }
\endinsert

\subsection{Scalars}

Our first class of models simply consists of complicating the SM Higgs sector
by adding various scalar multiplets to the standard $Y=\hf$ doublet. Since, at
one loop, each new multiplet contributes additively to the oblique parameters,
we may consider the contributions of such new particles one multiplet at a
time.

\ref\lavli{L. Lavoura and L.F. Li, \prd{49}{94}{1409}.}
\ref\evans{N. Evans, \prd{49}{94}{4785}.}

\figure\scalarfigone{The contribution of various scalar multiplets to the
oblique
electroweak parameters $S'$ and $T'$. Solid line: a colour-singlet, $Y=1$
triplet
with masses $(m_2^2,m_1^2,m_0^2) = (a^2,\hf(a^2+b^2),b^2)$, where $a$ and $b$
take the values given in brackets: $(a,b)$. Dotted line: A colour-singlet,
$Y={3 \over 2}$ doublet with masses $(m_2,m_1)$ as indicated in brackets.
The grid in both cases represents steps of 30 GeV.}

\figure\scalarfigtwo{More scalar-generated oblique parameters.
Solid line: a colour-triplet, $Y=\nth{6}$ doublet (squarks)
with masses $(m_{\tilde{u}},m_{\tilde{d}})$. Dotted line: a colour-singlet,
$Y=\hf$ doublet (a new Higgs or slepton \etc) with masses $(m_1,m_0)$.
The grid indicates steps of 25 (resp. 30) GeV for the solid (resp. dotted)
plots.}

\figure\scalarfigthree{The oblique corrections due to a colour-singlet $Y=0$
real triplet with masses $(m_0,m_1)$ given in brackets. The grid spacing is 30
GeV.}

\figure\scalarfigfour{More oblique parameters from scalar loops. Dotted line: a
colour-singlet $Y=0$ doublet with masses $(m_{\hf},m_{-\hf})$. Solid line: a
colour-singlet, $Y=0$ real doublet plotted against its mass, $m_{\hf}$.
The grid spacing is 30 (resp. 10) GeV for dotted (resp. solid) plots.
Small Figure: Here the lower line reproduces the colour-singlet, $Y=0$ real
doublet as above, while the upper line gives a colour-singlet, $Y=1$ singlet
having mass $(m_1)$. The Grid spacing in both cases is 10 GeV.}

\ref\lepbounds{Review of Particle Properties ,\prd{45}{92}{}.}

The contributions of scalar multiplets to the parameters $S$ -- $X$ have been
computed in Refs. \lavli\ and \evans.\foot\nonminexample
{In fact, it is amusing that one of the cases worked out in Ref. \lavli\
furnishes an example of a scalar multiplet for which $\ss S'$ and $\ss T' $
take
the central values of the fit of Ref. \peskin.  The model which does so is the
eleven-dimensional multiplet  having weak isospin $\ss J = 5$ and weak
hypercharge $\ss Y = -\hf$, and for which the eleven states are equally split
in
squared mass, starting with the lowest-mass state at $\ss m_1 =134$ GeV and the
highest-mass state at  $\ss m_2 = 159$ GeV. With these choices one finds $\ss
S'
= -0.58$ and $\ss T' = -0.38$.}
We have surveyed the parameter space of couplings and masses for various
scalar multiplets, searching for the regions which can contribute negatively
to $S'$ and $T'$. We present the results of this survey in two different ways.
First, Figs.~\scalarfigone\ through \scalarfigfour\  illustrate the dependence
of the oblique parameters
on scalar masses, by displaying the region of the $S' - T'$ plane which can be
reached by varying the scalar masses from 50 to 200 GeV. Next, we display in
Table I the values for these masses which are `optimal', meaning that they
maximize the magnitude of the contribution to $S'$ and $T'$ subject to the
condition that their ratio satisfies $S'/T' = (0.58/0.38)$. We choose this
ratio
as indicating the direction in the  $S' - T'$ plane of the central value of the
fit of Ref. \peskin. By doing so, we do not intend to argue that this ratio has
been definitively determined by the data, but rather to give a quantitative
indication of the size that is possible for the oblique parameters in each
case.
In searching for the masses which are optimal in this sense, we never permit
any
of our scalar masses to fall below 50 GeV, to avoid the bounds from direct
production at LEP \lepbounds. Unless  stated otherwise,
all of the numbers assume the new multiplets are colour singlets.

\ref\zeemodel{A. Zee, \plb{161}{85}{141}.}
\ref\squarkbounds{The CDF Collaboration, F. Abe {\it et al.},
\prl{69}{92}{3439}.}

We consider the following types of scalar multiplets:
\topic{Isosinglet Scalars}
The simplest possible scalar multiplet to add is an \sutwo\ singlet. Such
particles arise in several interesting theoretical scenarios. They arise:
($i$) as scalar partners to the right-handed leptons and quarks in
supersymmetric models; ($ii$) in the class of models proposed by Zee \zeemodel\
some years  ago; and ($iii$) in leptoquark models
where they can couple leptons to quarks in unorthodox, but baryon- and
lepton-number preserving, ways.
\topic{Isodoublet Scalars}
\sutwo\ doublets form a particularly well-explored wrinkle to the fabric of
the minimal standard model, since a second $Y= \hf$ scalar doublet
appears naturally in many of its alternatives. Among the models which
naturally incorporate doublet scalars are: ($i$) supersymmetric models,
for which the extra scalars arise as an additional Higgs doublet (with $Y =
\hf$),  as well as the scalar superpartners of the left-handed quarks and
leptons (having $Y=\nth{6}$ and $\hf$ respectively); and ($ii$) models of
(spontaneous or explicit) $CP$-violation at the electroweak scale, such as
might be required for electroweak baryogenesis. Since the superpartners
of the  left-handed quarks are subject to stringent CDF bounds \squarkbounds\
they cannot contribute significantly to negative $(S',T')$, as can be
seen from Fig. \scalarfigtwo . For this reason
they are omitted from Table I.
\topic{Isotriplet Scalars}
Isotriplet scalars arise in many situations, such as in left-right symmetric
models. Typically, if these fields are permitted to acquire vacuum expectation
values ({\it vev}'s),  they can spell trouble for low-energy weak-interaction
measurements, through their tree level contributions to the rho parameter (\ie\
$T$). We sidestep these bounds by assuming all {\it vev}'s to be zero.
\endtopic

Several features emerge from an inspection of Table I and Figs.~\scalarfigone\
through \scalarfigfour.
\topic{1}
All of the allowed regions in these figures include the origin, $S' = T'
=0$, with the corollary that it is always possible to produce negative values
for  $S'$ and $T'$. There is a simple explanation why this is always so for
scalars, even though, as we shall see, the same is {\it not} true for fermions.
The main point is that all of the oblique electroweak parameters must vanish in
the limit that the new physics becomes heavy in an \gwk-invariant way. And,
unlike for fermions, gauge-invariant masses are {\it always} possible for
scalar
multiplets.
\topic{2}
We generally find the largest values for $S'$ and $T'$ arising from the
smallest values for the scalar masses. As a result, the most important regions
of parameter space are precisely those for which the difference between $S$ and
$S'$ and $T$ and $T'$ is the most important.
\topic{3}
Although $S'$ frequently becomes more negative for large splittings
within a scalar multiplet, the growth of $T$ in this limit invariably
drives $T'$ positive, thereby forcing a preference for roughly equal masses
within the multiplet.
\topic{4}
Typically, larger values for $S'$ and $T'$ are possible given larger values
for $Y$, or given a larger \sutwo\ representation. This can be most clearly
seen
from Table I, for which $S'$ increases both with increasing weak isospin,
and with increasing $Y$ for fixed weak isospin. The variation with $Y$
is simplest to see for \sutwo\ singlets, for which all of the oblique
parameters
are simply proportional to $Y^2$.
\topic{5}
The oblique parameters are similarly enhanced if the scalar multiplets couple
to the strong gauge group, \suthree. In this case all oblique parameters must
be multiplied by the dimension, $d_c$, of the appropriate \suthree\
representation. (These factors, together with the factors of hypercharge
mentioned earlier, can be quite large. For instance, a colour-sextet
\sutwo-singlet scalar having $Y={4 \over 3}$, as would be required for a Yukawa
coupling to two right-handed up-type quarks, has $d_c Y^2 = {96 \over 9} = 10
\,{2 \over 3}$.)

\ref\tevatronbounds{The CDF Collaboration, F. Abe \etal, \prd{48}{93}{3939}.}
\ref\herabounds{The H1 Collaboration, I. Abt \etal, \npb{396}{93}{3}; The ZEUS
Collaboration, M. Derrick \etal, \plb{306}{93}{173}.}

Using colour to amplify the oblique corrections does not come without its
price,
however, since the masses of coloured particles are often subject to more
stringent bounds than are those for colour singlets, due to their
non-observation in
$p-\overline{p}$ \tevatronbounds\ and $e - p$ \herabounds\ collisions.  These
bounds can be sensitive to the nature of the new particle's dominant decay
mode, however, and so can be more model dependent than are those furnished
by LEP.
\topic{6}
The one-loop contributions to oblique parameters are fairly robust,
depending solely on the assumed electroweak transformation properties of
the multiplet (in the absence of significant mixing amongst various
electroweak multiplets). The same is generally not true for the other bounds on
new  scalar multiplets, since these can depend on such things as the existence
and  strength of their Yukawa couplings to fermions, as well as on whether or
not they acquire nonzero {\it vev}'s. In fact,
it is  quite simple to arrange for such particles to contribute to experiment
predominantly through their oblique corrections, just by forbidding (or
suppressing) their non-gauge couplings by a (possibly approximate) symmetry.
\topic{7}
Finally, it is clear that provided no particle masses are permitted to fall
below 50 GeV, no single scalar particle can by itself account for a large
negative
value for  $S'$ and $T'$. Should spinless particles be required to explain a
trend to negative $S'$ and $T'$, if this were to persist as the data improves,
it would require a number of new scalars, all contributing together to
produce the desired-size effect. As we see next, the same need not be true for
exotic fermions, whose contributions to oblique parameters can be considerably
larger.
\endtopic

\subsection{Fermions}

We now turn to the addition of exotic fermions of various types to the Standard
Model. As we found for scalars, at one loop we may consider separately the
contribution to the oblique parameters of each additional multiplet. A
calculation of the six oblique parameters as functions of general fermion
masses
and couplings is given in Ref. \alphabet.

\figure\fermionfigone{The contribution of an extra
SM family to the oblique parameters. Solid line: a colour-triplet, $Y=\nth{6}$
quark doublet with masses $(m_{b'},m_{t'})$ as indicated in brackets. Dotted
line:
a colour-singlet, $Y=\hf$ lepton doublet with masses $(m_{\nu '},m_{l'})$.
The grid spacing represents steps of 25 (resp. 30) GeV for the solid (resp.
dotted)
plots.}

\figure\fermionfigtwo{Oblique parameters due to colour-singlet
mirror fermions. Solid line: a $Y=1$ mirror singlet with mass
$(m_1)$. Dotted line: a $Y=0$ mirror doublet with (degenerate) mass
$(m_\hf)$. Dash-dotted line: a, $Y=\hf$ mirror doublet with (degenerate)
mass $m_1=m_0$. The grid spacing represents steps of 10 GeV for
all three plots.}

\midinsert

$$\vbox{\tabskip=0pt \offinterlineskip
\halign to \hsize{\strut#& #\tabskip 1em plus 2em minus .5em
&\hfil#\hfil &#&\hfil#\hfil &#&\hfil#\hfil &#&\hfil#\hfil
&#&\hfil#\hfil &#&\hfil#\hfil &#&\hfil#\hfil &#& \hfil#\hfil
&# \tabskip=0pt\cr
\noalign{\hrule}\noalign{\smallskip}\noalign{\hrule}\noalign{\medskip}
&& Multiplet && Optimal Masses && $S'$ && $T'$ && $U'$ && $S$
&& $T$ && $U$ &\cr
&& $(3 \times 2 \times 1)$ && (GeV) &&  &&  &&  &&  &&  && &\cr
\noalign{\medskip}\noalign{\hrule}\noalign{\medskip}
&& (\bfone,\bftwo,$Y=-\hf$)${}^\star$ && $\pmatrix{m_{\nu'} \cr m_{\ell'}
\cr} = \pmatrix{ 50 \cr    92\cr}$ && -0.2 && -0.1  && 0.1 && -0.06 && 0.03 &&
0.04 &\cr
\noalign{\smallskip}
&& (\bfthree,\bftwo,$Y=\nth{6}$)${}^\star$ && $\pmatrix{m_{t'} \cr m_{b'} \cr}
=
\pmatrix{104 \cr 50 \cr}$  && -0.1 && -0.08  && 0.3 && -0.01 && 0.2 &&
0.1 &\cr
&&  &&  &&  && && &&  &&  && &\cr
&& (extra SM family) && $\pmatrix{m_{\nu'} \cr m_{\ell'} \cr m_{t'} \cr
m_{b'} \cr} = \pmatrix{50 \cr 100 \cr 130 \cr 95 \cr}$
&& -0.06 && -0.04  && 0.2 && 0.04 && 0.1 && 0.05 &\cr
&&  &&  &&  && && &&  &&  && &\cr
&& (\bfone,\bfone,$Y=1$)${}^\dagger$ && $m = 50$ && -0.2 && -0.09 && 0.03 &&
-0.03 && 0 && 0.03 &\cr
&& (\bfone,\bftwo,$Y=0$)${}^\dagger$ && $m = 50$ && -0.4 && -0.5  && 0.4 &&
-0.2
&& 0 && 0.009 &\cr
\noalign{\smallskip}
&& (\bfone,\bftwo,$Y=\hf$)${}^\dagger$ && $m = 50$ && -0.5 && -0.5  && 0.4 &&
-0.2 && 0 &&0.005 &\cr
\noalign{\medskip}\noalign{\hrule}\noalign{\smallskip}\noalign{\hrule} }}$$
${}^\dagger$ {\eightrm Plus a mirror multiplet with conjugate quantum numbers.}
\hfil\break
${}^\star$ {\eightrm Plus right-handed isosinglets with identical electric
charges.}
\smallskip

\centerline{\bf Table II: Exotic Fermions}
\medskip
\noindent {\eightrm One-loop oblique electroweak parameters due to exotic
fermions.  This table displays the masses which `optimize' the oblique
electroweak parameters in the sense described in the text, together with the
resulting optimal values. $\ss ({\bf r_1}, {\bf r_2}, Y=y)$ denotes the
transformation properties of the left-handed fermions.}

\endinsert

\ref\familybounds{F. Abe {\it et al.}, \prl{64}{90}{147}, \prl{68}
{92}{447}. See also J.F. Gunion, D.W. McKay and H.Pois, preprint
UCD-94-25 (hep-ph/9406249) and references therein.}

We present our survey of the parameter space of couplings and masses for
the various fermion multiplets in the same way as we did for the exotic
scalars:
Figs.~\fermionfigone\ and \fermionfigtwo\ plot the dependence of the oblique
parameters on the
fermion masses, which we take to range from 50 to 200 GeV. Table II displays
the values for these masses which are `optimal' in the same sense as was used
for the scalars.  While optimizing we forbid masses which are smaller
than 50 GeV, due to the bounds from direct production at LEP \lepbounds. An
exception is the case of an extra family, where Tevatron bounds \familybounds\
on $4^{th}$ generation quarks are much stronger than are those from LEP.
It is possible, should the $4^{th}$ generation quarks not mix with the usual
ones,
that the Tevatron bounds would be somewhat weaker.
For completeness we therefore include the optimized values for an extra quark
doublet, subject only to the LEP bounds, in Table II.
All exotic fermions are taken to be colour
singlets, except for fourth-generation quarks which are taken to be triplets.

The following types of fermion multiplets are of particular interest:
\topic{Isosinglet Fermions}
Once more the simplest possible addition is an \sutwo\ singlet.  We assume
here a mirror fermion for which the left- and right-handed hypercharge are
equal, so that arbitrary masses are possible without breaking \gwk\ invariance.
\topic{Isodoublet Fermions}
There are two kinds of isodoublet fermions which have been widely considered
in the literature. These are ($i$) sequential quarks and leptons, as would be
found in a fourth generation of SM particles, for example, or ($ii$) \sutwo\
doublets of mirror fermions for which the left- and right-handed parts are both
doublets with identical hypercharge quantum numbers.
\endtopic

\ref\technideath{M.E. Peskin and T. Takeuchi, {\it ibid.}; W.J. Marciano and
J.L. Rosner, \prl{65}{90}{2963}; D.C. Kennedy and P. Langacker,
\prl{65}{90}{2967}.}

Several points concerning Figs.~\fermionfigone\ and \fermionfigtwo\
and Table II deserve emphasis.
\topic{1}
Unlike for the scalar case, it is not necessarily true that the origin, $S' =
T'
=0$, need lie in the parameter space of an exotic fermion multiplet, provided
that this multiplet lies in a chiral representation of \gwk\ (as does an
additional sequential lepton or quark). A quark multiplet, in particular,
definitely favours positive $S'$, making a real trend to negative $S'$ strongly
disfavor such  particles. (This is essentially the observation which
was originally used to
disfavor technicolour models \technideath.) As may be seen from
Fig.~\fermionfigone\ or Table II, for a complete additional generation the
contribution of extra quarks can be compensated by the additional leptons, but
{\it only} if the additional leptons are reasonably light.
\topic{2}
As is the case for scalars, the largest values for $S'$ and $T'$ arise from
the smallest values for the fermion masses; again emphasizing the difference
between the primed and unprimed oblique parameters.
\topic{3}
We reproduce here the growth of $T$, and hence the low-energy rho parameter,
in the limit that the mass splitting in a standard multiplet becomes large.
\topic{4}
Fig.~\fermionfigtwo\ and Table II display the values for
$S'$ and $T'$ that are obtained
for a doublet of mirror fermions having various hypercharges. Only the case
of a degenerate multiplet is shown here because it is only in this case that
all
of the parameters $S$ through $X$ are independent of the renormalization
scale, $\mu$. (This is in contrast with all of the previous examples we
consider,
which are $\mu$-independent for any choices for the masses.) For a
nondegenerate
mirror fermion multiplet the parameter $T$ develops a $\mu$-dependence
which is proportional to the square of the mass splittings within the
multiplet.

There is a simple reason for the appearance of the $\mu$ dependence for a
nondegenerate multiplet. The main point is that although a mirror doublet
can acquire a common degenerate mass in an \gwk-invariant way, renormalizable
interactions can split the masses within a multiplet only if new, non-doublet,
scalars
are introduced and acquire a {\it vev}. As a result these scalars contribute to
the parameter $T$ at the tree level, and so this parameter must be renormalized
---
thereby developing a dependence on $\mu$ --- just as must any other classical
parameter.
\topic{5}
The magnitude of $S'$ and $T'$ grows with the hypercharge and size of the
colour representation for the multiplet concerned. Notice, however, that
the overall contribution of a
given fermion representation is significantly larger than that of additional
scalars which transform in the same representation.
\endtopic


\section{Conclusions}

We have explored in this note the kinds of new physics which can produce
deviations from the SM predictions for $Z$-pole physics in the direction of
negative values for the Peskin-Takeuchi-like parameters $S'$ and $T'$. We have
been motivated to do so by the suggestive --- if presently statistically
inconclusive --- reinforcement of the trend in this direction found by fits
to these oblique parameters which combine the most recent $\alr$ measurement
with those at LEP.

We have found it to be reasonably easy to construct models for which the
contributions of new physics to precision electroweak measurements are well
approximated by purely oblique vacuum-polarization effects. It is more
difficult, but not impossible, to obtain a correction which predicts $S'$
and $T'$ as large as $-0.1$ to $-0.6$. New fermions are preferable
to new scalars in this regard, since they generate vacuum polarizations which
are systematically larger than the scalars, given similar couplings and masses.

In all cases we find that requiring relatively large contributions to
the oblique parameters points to new particles whose masses are {\it not} very
large compared to $\mw$ or $\mz$. We emphasize the necessity for interpreting
the data on the $Z$ pole (and the $W$ mass) in this case in terms of the
variables $S',T'$ and $U'$, which are linear combinations of the usual
variables, $S, T, U$ with the new ones, $V$ and $X$, of Ref.~\alphabet. In
fact, we find that the difference between the primed and unprimed parameters is
important for allowing a larger region of parameter space to contribute
acceptably to the electroweak parameters in these models.

This preference for comparatively light particles should have happy
consequences should the central values continue to prefer negative $S'$ and
$T'$ as the accuracy of the data improves. Since the new particles which we
consider are comparatively light, they stand a good chance of being seen in
other experiments once higher energies become directly observable. In
particular, since it is the coupling of these particles to the $W$ and the $Z$
which is responsible for their contributions to the oblique vacuum
polarizations, they should be directly pair-produced at LEP-200 if they are
light enough for this to be kinematically allowed. Their effects can also be
searched for in
electroweak measurements at lower energies, $q^2 \simeq 0$, since
their contributions to the oblique parameter $X$ of Ref.~\alphabet\
causes the value of $\sin\theta_w$ as measured at low energies to differ from
that measured at $q^2 = \mz^2$. Similarly, contributions to the parameter
$W$ give rise to deviations of the $W$ width from its value as predicted by
the SM supplemented by $S$, $T$ and $U$.
Signals at HERA or the TeVatron could
also be expected (or not) depending on the more detailed features of the new
particles' masses and couplings.

Interestingly our analysis tends to disfavor an additional $SM$ family of
fermions,
provided that it mixes with the usual three families, and that the lepton
masses
(especially the neutrino mass) are not too close to their present LEP lower
limits
$\simeq M_Z/2$. This reasoning goes along the same lines as those used to
disfavor technicolor models \technideath .

However, it is clearly premature to be discarding models based on the size of
their
negative contributions to $S'$ and $T'$. Our intention here is not to do this,
we merely wish to determine what kinds of new-physics candidates can produce
oblique corrections that are qualitatively in the right direction, should the
trend to negative values for $S'$ and $T'$ ultimately become statistically
significant. If this happens, we hope that our results for the magnitude of
these parameters (and $U'$) as functions of the assumed type of new physics
will make a useful starting point for more detailed investigations.

\bigskip
\centerline{\bf Acknowledgments}
\bigskip

We would like to thank Ken Ragan for helpful conversations,
and Tom Rizzo for giving us access to the unpublished
results of various recent fits to the data.
This research was partially funded by funds from the N.S.E.R.C.\ of Canada,
les Fonds F.C.A.R.\ du Qu\'ebec, and the Swiss National Foundation.

\listrefs

\figurecaptions

\bye